\documentclass[useAMS]{mn2e}
\usepackage{graphicx}
\newcommand{\quotes}[1]{``#1''}
\def\ltsima{$\; \buildrel < \over \sim \;$}
\def\gtsima{$\; \buildrel > \over \sim \;$}
\def\lsim{\lower.5ex\hbox{\ltsima}}
\def\gsim{\lower.5ex\hbox{\gtsima}}

\title[The Horizontal Branch of M3, M13 and M79.]
       {The Horizontal Branch in the UV colour Magnitude Diagrams. II. The case of M3, M13 and M79.}
\author[Dalessandro et al.]
  {E.~Dalessandro,$^1$ M.~Salaris,$^2$ F.R.~Ferraro,$^1$
  A.~Mucciarelli,$^1$ S.~Cassisi,$^3$  \\
    $^1$Dipartimento di Fisica e Astronomia, Universit\`a degli Studi
 di Bologna, viale Berti Pichat 6/2, I--40127 Bologna, Italy\\
 $^2$Astrophysics Research Institute, Liverpool John Moores University,
 Twelve Quays House, Egerton Wharf, Birkenhead CH41 1LD, UK\\  
 $^3$INAF - Osservatorio Astronomico di Collurania, via Mentore Maggini, 
 64100 Teramo, Italy
}

\date{18 December, 2012}

\pagerange{\pageref{firstpage}--\pageref{lastpage}} \pubyear{2012}

\def\LaTeX{L\kern-.36em\raise.3ex\hbox{a}\kern-.15em
    T\kern-.1667em\lower.7ex\hbox{E}\kern-.125emX}

\begin{document} 

\label{firstpage}

\maketitle
\begin{abstract}

We present a detailed comparison between far-UV/optical colour Magnitude Diagrams obtained with 
high-resolution {\it Hubble Space Telescope} data  
and suitable theoretical models for three Galactic Globular Clusters: M3, M13 and M79.\\
These systems represents a \quotes{classical} example of clusters in the intermediate metallicity 
regime that, even sharing similar metal content and age, show remarkably different 
Horizontal Branch morphologies. As a consequence, the observed differences in the colour
distributions of Horizontal Branch stars 
cannot be interpreted in terms of either first (metallicity) or a second parameter such as age.
\\
We investigate here the possible role of variations of initial Helium abundance ($Y$). 
Thanks to the use of a proper setup of far-UV filters, we are able to put strong constraints on the 
maximum Y ($Y_{max}$) values compatible with the data. 
We find differences $\Delta Y_{max}\sim 0.02-0.04$ between the clusters with M13 showing the largest
value ($Y_{max}\sim0.30$)
and M3 the smallest ($Y_{max}\sim0.27$). In general we observe that these values are correlated with 
the colour extensions of their Horizontal Branches and with the range of the observed Na-O anti-correlations.

\end{abstract} 


\section{Introduction}																	   
Horizontal branch (HB) stars are the progeny of low-mass Red Giant Branch stars (RGB) 
burning helium in their cores and hydrogen in a shell around it (Hoyle \& Schwarzschild 1955).
As first noticed by Iben \& Rood (1970),
the different HB star colour distributions observed in old stellar systems, is the 
reflection of the amount of mass lost during the RGB phase.\\
The scientific community agrees from nearly fifty years about the fact that the
principal parameter governing the shape of HBs in Galactic 
Globular Clusters (GGCs) is metallicity. The general rule is that
metal-rich systems have red HBs, while in the metal-poor ones stars are 
distributed on average at higher
effective temperatures (bluer colours). Several exceptions have come out 
during the last decades; remarkable cases
the cases of NGC6388 and NGC6441 (Rich et al. 1997), which despite their
metallicity ($[Fe/H]\sim-0.6$) show some of the bluest HBs known among GGCs
 (Busso et al. 2007; Dalessandro et al. 2008).
Moreover several clusters, sharing similar metal content, reveal different 
HB morphologies, typical cases being the pairs  
NGC5927 - NGC6388 at high metallicities
($[Fe/H]\sim-0.4$),
M3 - M13 at intermediate 
metallicity regime ($[Fe/H]\sim-1.5$; Ferraro et al. 1997) and M15 - M92 at low metallicities
($[Fe/H]\sim-2.3$).\\
These noticeable exceptions have required 
the introduction of a 
second (Freeman \& Norris 1981) and possibly a third parameter in order to explain the HB distributions
in all GGCs.
What we can call now the \emph{\quotes{i-th parameter problem}} 
is still a hot topic, as stressed by several authors, we recall the reader to Catelan 2009 for a nice
review (see also Dotter et al. 2010 and Gratton et al. 2010; hereafter D10 and G10 respectively).\footnote{Moreover the increase of photometric capabilities has allowed to reveal tiny
differences and features like gaps along the HB of some GGCs 
(see Ferraro et al 1998 for example), has complicated the interpretation.} \\
An accurate knowledge of the physical parameters playing a role in shaping the 
HB is extremely important also 
for an appropriate interpretation of distant unresolved stellar populations.
In fact it is well known that the HB morphology can have a strong impact on the
integrated light of stellar populations, affecting colours and line indices
(Lee et al. 2002; Schiavon et al. 2004; Percival \& Salaris 2011; Dalessandro et al. 2012).\\
Despite the huge efforts made to address this problem, its solution 
is not obvious and still different
scenarios are proposed. One of the reasons that complicates the identification of
the mechanisms -- other than metallicity -- at work in shaping the 
observed luminosity and effective temperature distribution of stars along the HB 
is that there are many possible culprits (mass-loss, age, helium abundance ...; 
see Rood 1973 for example) 
and some of them are not well constrained from theory.\\
Age has been identified as the natural global second parameter by many authors in the past years (Lee et al. 1987, 1988, 1990;
Lee, Demarque \& Zinn 1994; Sarajedini \& King 1989).  According to this
interpretation older clusters tend to have bluer HBs, while younger ones should have on average redder
HB morphologies.
This scenario appeared in agreement with the picture for the Galaxy formation and its early evolution
(Searle \& Zinn 1978; Zinn 1985). 
By means of high resolution HST data for a large sample of GGCs,
D10 found that the existence of outer halo GCs with 
anomalously red HBs fits well the scenario in which age is the second parameter.
In fact, the behaviour of the 4-5 relatively younger clusters in their sample
could be reproduced in term of correlation between age and HB morphology, 
while the bulk of the analyzed 
targets is peaked around old ages (see ages reported by Salaris \& Weiss 2002, G10, D10)
and doesn't show  any obvious correlation. 
Also results by G10 agree on the fact that age is the second main parameter driving the  HB 
morphology. \\
It is also worth noticing that most of these results are based on 
optical CMDs and HB morphology parameters (like the well known HBR from Lee et al. 1994), which 
tend to minimize the importance of blue tails. 
On the contrary using proper combinations of Ultra-Violet (UV) and optical filters
has an important impact both in term of HB classification and comparison with theoretical models.

Still, age is not able to explain exhaustively the HB morphology. 
Detailed cluster to cluster comparisons 
have shown that there are systems with similar iron content and age, but 
remarkably different HB morphologies. A clear example is given by the three clusters 
M3 - M13 - M80, 
as shown by Ferraro et al. (1997, 1998) and at present there is hardly a 
scenario able to give a satisfactory explanation for their different morphologies.\\    
As suggested by Buonanno et al. (1985) and Fusi Pecci et al. (1993), age might be one of many 
and probably the most important
HB second-parameter, but not the only one. Buonanno et al. (1995) argued that it is not possible
to reproduce the complex HB zoology with a single \quotes{global} parameter, but more likely we can
explain it by thinking of a \quotes{global combination} of \quotes{non-global} quantities and phenomena 
related to the formation, chemical and dynamical evolution of each cluster.\\
The necessity of at least a third parameter transpires also from D10 and G10 analyses, 
in the form of either the luminosity cluster density or stellar density ($\log(\rho)$) 
-- as already suggested by Fusi Pecci et al. (1993) -- 
which might correlate with the hot extension of the HBs, or 
a variation of the initial helium abundance (Y), respectively.\\
Indeed D'Antona et al. (2005) and Dalessandro et al. (2011 -- hereafter PaperI) have shown 
that for NGC2808 the main parameter that determines the HB morphology 
is Y.
In particular in PaperI we have been able to satisfactory reproduce the cluster complex HB morphology by assuming 
three different
sub-populations with He abundances compatible with what inferred from the multimodal Main Sequence (MS; Piotto et al. 2007)
and spectroscopic analyses (Bragaglia et al. 2010; Pasquini et al. 2011). It
is worth noticing however that only few clusters 
show evidences of multiple sub-populations with different He abundances as inferred from their MS or spectroscopic
analyses (NGC2808, 
$\omega$ Centauri -- Bedin et al. 2004, NGC6752 -- Milone et al. 2010, NGC6397 -- Milone et al. 2012; NGC1851 -- Gratton et
al. 2012) \\
In order to further investigate the \emph{\quotes{i-th parameter problem}}, 
we selected a triplet of GGCs with similar 
metallicity ($[Fe/H]\sim-1.50$) and age, namely 
NGC5272 (M3), NGC6205 (M13) and NGC1904 (M79) as templates for a comparative
analysis of their HBs. These targets are basically the same used by Ferraro et al. (1997,
1998), with the only exception of M80, which have been replaced by M79 since recent
spectroscopic analyses (Carretta et al. 2009) show that it is slightly more metal-poor
than the others.  
As shown by Ferraro et al. (1998), M13 has a very extended HB with 
two clear gaps (named G1 and G2 after their analysis);  
while, the HB of M3 covers a much narrower extension in 
effective temperature and shows only a mild indication for a gap at the same $T_{\rm eff}$
as G1 in M13. M79 represents an intermediate case between the two. \\
As already done for NGC2808 (Paper~I) we will use a combination of optical and
UV photometric images. Details about data reduction are presented 
in Section 2. In Section 3 we will focus on the long-standing debate about the age-differences between
these clusters. Comparisons with
theoretical models and cluster-to-cluster differences will be discussed in Section 4 and 5.

\begin{figure}
\includegraphics[width=84mm]{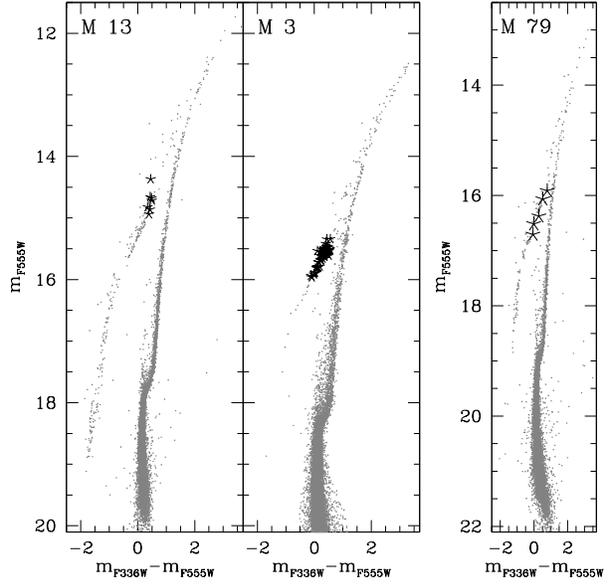}
\caption{Optical CMDs of the three GGCs considered in our analysis.}
\label{cmdopt}
\end{figure} 
 
\section{Observational data sets and data reduction}

The data-set used in the present work consists of Hubble Space Telescope (HST)
Wide Field Planetary Camera 2 (WFPC2) images. All targets
have been observed in the $F555W$, $F336W$ and $F160BW$ bands, the only exception is M3 
for which no F160BW data are available.
(see Table~1). 
Images and photometry obtained in other bands 
are also available for some of the clusters, however 
as shown in Rood et al. 2008 and Paper~I
the combination of $F555W$, $F336W$ and $F160BW$ seems to be the best
filters setup to study both RHBs and stars lying at the very hot end of the HB.
The photometric catalogues of M3 and M13 have been already 
partially presented by Ferraro et al. (1998), while the photometry of M79  
has been already used by Lanzoni et al. (2007). Moreover Far-UV photometry 
of M79 and M13 has been presented by Rood et al. (2008) in order 
to describe in detail the advantages of using the combination of $F160BW$ and $F555W$
magnitudes for comparison with HB models.\\
\\

\begin{figure}
\includegraphics[width=84mm]{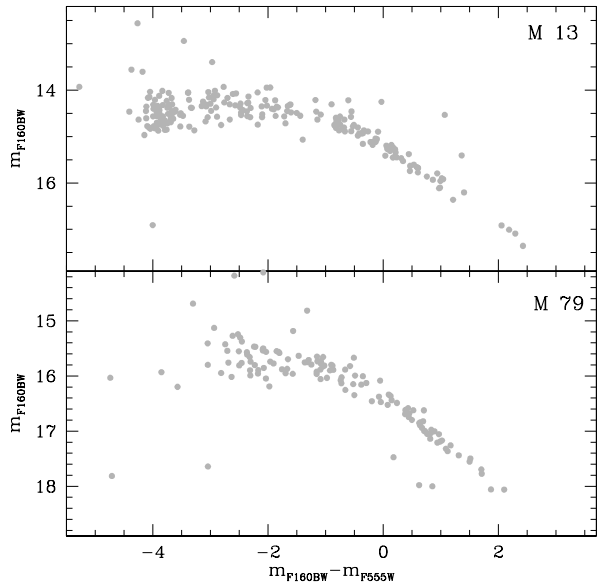}
\caption{Far-UV CMDs of M13 and M79.}
\label{cmdfuv}
\end{figure}

In all cases the photometric analysis has been performed by using
ROMAFOT (Buonanno et al. 1983). The reduction strategy has been 
optimized to be particularly sensitive to hot stars, as already
described by Ferraro et al. (2003). For the present analysis 
attention has been paid to the cross-identification of stars
detected in the $F160BW$ images. 
Stars observed only in these bands have been force-fitted 
in the other bands (see Paper I).  \\
The instrumental magnitudes have been calibrated to the VEGAMAG photometric
system by using zero-points and the gain settings reported in 
Table 5.1 of the WFPC2 
\emph{data handbook Manual}\footnote{http://www.stsci.edu/documents/dhb/web/}.
Magnitudes thus obtained have been also corrected for Charge 
Transfer Efficiency (CTE) effect by means of the prescriptions
by Dolphin (2000) and updated equations listed in the dedicated page of 
Dolphin's web site\footnote{http://purcell.as.arizona.edu/}.\\
The resulting colour Magnitude Diagrams (CMDs) for each cluster are shown
in both the 
($m_{F160BW}$, $m_{F160BW} -m_{F555W}$) and
($m_{F336W}$, $m_{F336W}-m_{F555W}$) planes in Figs~\ref{cmdopt} and ~\ref{cmdfuv}.

Star lists have been reported to the absolute coordinate system 
by mean of proper roto-traslations. In particular, for each cluster 
the instrumental coordinates have been reported to the GSCII astrometric 
standard system by using 
a large number of primary and secondary standards in common,
consisting typically 
of Wide Field ground-based data-sets,
in the WFPC2 Fields of View (FOVs) (see 
Lanzoni et al. 2007 for details). At the end of the procedure,
the typical error of the astrometric solution is $\sim 0.2\arcsec$.

\subsection{RR Lyrae identification}

The identification of variable stars lying along the HB is a crucial
step to perform a suitable comparison between observations and theoretical models. This is particularly
true in case of M3, that hosts tens of RR Lyrae stars.\\
However, none of the data-sets used here is designed for variability search,
both because of the short time baseline covered and number of images acquired.
In our CMDs, variables are basically sampled at random phase.
Thus we decided to use publicly available
catalogues of variable stars in order to cross-identify them in our photometric catalogues,
as already done in Paper I. We retrieved position, magnitude, and classification lists
from the \quotes{Catalogue of Variable Stars in Galactic Globular Clusters
(2011)}\footnote{http://www.astro.utoronto.ca/~cclement/read.html} (Clement et al. 2001).
We refer to that paper for all details and references.\\
We focused only on RRLyrae stars, neglecting all the other different variable types.
We first roto-traslated the catalogues of variables to our reference frames by using CataXcorr,
a software developed at the Bologna Observatory (Montegriffo, private communication).
We then looked for stars in common with our photometry. 
All the known RRLyrae stars in our FOVs have been recovered in our catalogues.
We found 59 RRLyrae in M3, 7 in
M13 and 5 in M79 (see Table~1). Their positions in the optical CMDs are shown in Figure~\ref{cmdopt}.

\section{Metallicity and relative ages}
An accurate determination of metallicities and ages is a crucial point for 
HB morphology studies and for the correct interpretation of comparative analyses.\\
The three GGCs in our sample have a similar iron content: according to 
the recent reassessment of the GC metallicity scale by Carretta et al~(2009), metallicities are equal to
[Fe/H]=$-1.50 \pm 0.05$, $-1.58 \pm 0.04$ and $-1.58 \pm 0.02$ for M3,
M13 and M79, respectively. Consistent results were also found by Sneden at al. (2004) who obtained 
[Fe/H]=$-1.55\pm0.02$ and $-1.57\pm0.07$ for M3 and M13 (for homogeneity we report only values obtained 
from the Lick Hamilton spectra). Only in the analysis by Cohen \& Melendez (2005) M3 and M13 are slightly 
more metal-rich, i.e. [Fe/H]=$-1.39\pm0.02$ and [Fe/H]=$-1.50\pm0.01$. In addition in this 
case the observed metallicity difference ($\Delta$[Fe/H]$\sim0.1$) is larger than the uncertainties.
\\

\begin{figure}
\includegraphics[width=98mm]{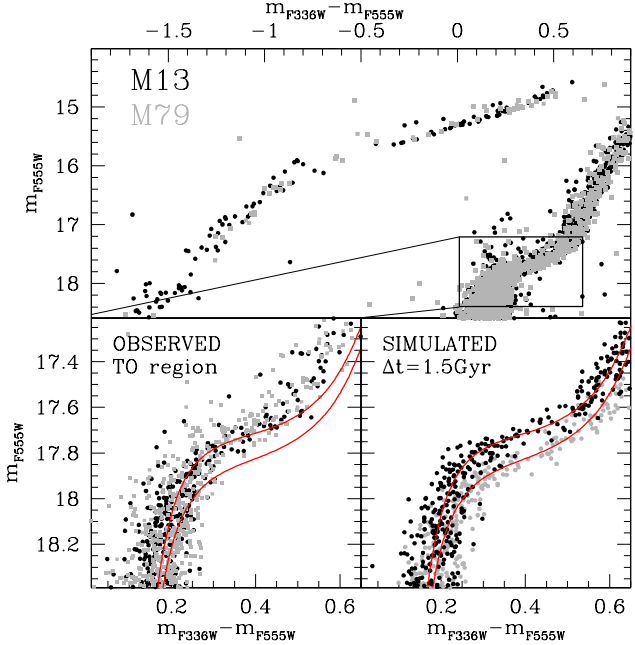}
\caption{Upper panel. Optical CMD of M79 super-imposed to M13 after correcting for 
differences in distance modulus and reddening. Bottom panels. Zoomed view of the TO region.
On the left, the observed CMDs are shown with two isochrones from the BaSTI database with $t=12$ Gyr 
and $t=13.5$ Gyr.
The right panel shows synthetic CMDs obtained from the adopted isochrones and the appropriate photometric 
errors.}
\label{m13m79}
\end{figure}

On the contrary the determination of their ages is a more debated argument. 
In literature there is a  quite general consensus that M3 and M79 are old and coeval GGCs.
There is instead more uncertainty about the age of M13.
Ferraro et al. (1997) put constraints on the possible 
difference in age between M13 and M3 by comparing their ridge mean lines 
obtained with the optical CMDs used in this work, and theoretical models by Dorman et al. (1996).
The age difference resulted to be smaller than $\Delta t=1-1.5$ Gyr. The same conclusions have 
been reached also by Johnson \& Bolte (1998) and Rosenberg et al. (1999).
According to the analysis by Grundahl et al. (1999) M13 may be younger than M3 by $0.7\pm0.2$ Gyr,
while for Rey et al. (2001) it is older by about $2$ Gyr.\\
Ages reported by other authors for the clusters of our analysis agree quite well within 
the error bars.
Salaris \& Weiss~(2002) found a typical difference
smaller than $\Delta t=1$ Gyr independently of the metallicity scale used.
Ages by D10 obtained using the isochrone-fitting method are basically identical;  
in fact t=$(12.5\pm0.5)$ Gyr and t=$(13.0\pm0.5)$ Gyr for M3 and M13 respectively,
while  M79 is not in their sample. The same results were obtained by Marin-Franch et al. (2009).
For G10 instead,  M13 is older than M3 and M79 of about $\Delta t =$ 1.75 Gyr.  \\
It is worth noticing that even moderate differences of the average Y among these three 
clusters do not affect appreciably the results of the methods employed to determine their relative ages 
(see, e.g., Marin-Franch et al.~2010). \\
We investigated any possible age difference by comparing the ($m_{F555W}, m_{F336W}-m_{F555W}$) CMDs of the three clusters of
our analysis.

\begin{figure}
\includegraphics[width=98mm]{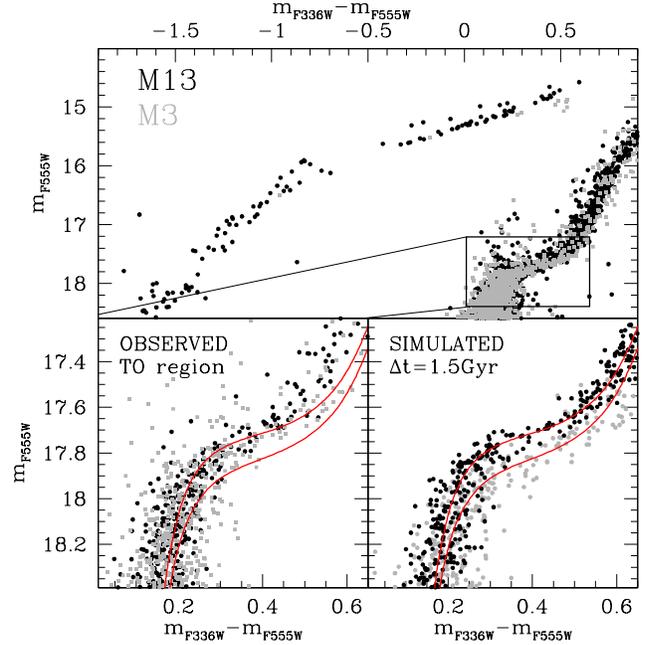}
\caption{As in Fig.~\ref{m13m79} but for M3 and M13.}
\label{m13m3}
\end{figure}
We used M13 as reference and shifted 
M79 and M3 according to the difference of $E(B-V)$ values and distance moduli
(see Sections 3.2, 3.3 and 3.4 for details about the values adopted).
As apparent from the upper panels of Figure~\ref{m13m79} and \ref{m13m3}, once relative
differences in reddening and distance are taken into account, the CMDs match well 
at the TO level and along the RGB, consistently with the fact that they have 
almost the same age and share a similar metallicity. \\
In the bottom panels  we focus on their Turn Off (TO) region. 
Also from a close inspection, the CMDs have been found to nicely overlap at the TO and SGB region. This already 
rules out any significant age difference between M13, M3 and M79.
However in order to perform a more accurate comparison, we also used two isochrones of proper metallicity 
and different age
from the BaSTI database (Pietrinferni et al. 2006). In particular 
we used as reference the isochrone that best fits the observed CMD ($t=12$ Gyr) and an additional one
that is $1.5$ Gyr older\footnote{Note that the aim of this analysis is note to obtain an estimate of the absolute age, 
but only to give constraints on relative ages}. As apparent from the left bottom panels of Figures \ref{m13m79} and \ref{m13m3},
the older isochrone does not match either the TO or the SGB.\\
Starting from these isochrones, and 
using the estimated photometric errors we build 
two synthetic populations of similar size to the observed ones for each comparison.
With the typical uncertainties of our HST catalogues at the TO magnitude level, 
an age difference of $1.5$ Gyr would be clearly visible and detectable. 
We then conclude that M13, M3 and M79 differ in age by less than $1-1.5$ Gyr.
This result is in agreement with estimates by several authors (Johnson \& Bolte 1998; Rosenberg et al. 1999;
Salaris and Weiss 2002;  
Marin-Franch et al. 2009; D10), while it is incompatible with the age differences proposed by G10.
It is worth noticing that an age difference of about $1.5-2$ Gyr would be compatible with the observed data
only assuming extremely different He abundances ($\Delta Y\sim0.1$) between these clusters.

\section{Theoretical framework}
\label{theory}

We have shown that M3, M13 and M79 have similar metallicities and are coeval within typical uncertainties.
Therefore in this case, the scenario proposed by G10 and D10, identifying   
the {\it second parameter} with age, is not applicable. We therefore investigate 
here the possible role of variations of Helium abundances in the framework of GCs having 
experienced multiple formation bursts (D'Ercole et al. 2008; Conroy \& Spergel 2011; Valcarce \& Catelan
2011)
 on a short timescale ($t\leq100$ Myr)
in environments enriched by material polluted by AGB stars (Ventura et al. 2002) or fast
rotating stars (Decressin et al. 2007). As highlighted before, Helium has been 
proposed also by G10 
to be the possible HB {\it third parameter}. \\
As in Paper~I, we have compared our photometric data with a suitable set of BaSTI\footnote{Available at the 
following URL: http://www.oa-teramo.inaf.it/BASTI.} $\alpha$-enhanced HB tracks 
(Pietrinferni et al.~2006). We have considered theoretical models for [Fe/H] =$-$1.62, 
which is the closest value available in the BASTI database 
to the clusters spectroscopically determined [Fe/H].

In our analysis we take advantage of all initial He-abundances available in the BaSTI database, 
i.e. Y=0.246, 0.300, 0.350 and 0.400. Bolometric corrections and 
extinction effects have been calculated as in Paper~I.

Recently, Gratton et al.~(2011) have found spectroscopical evidence of a moderate Na-O anti-correlation 
in stars belonging to the HB of NGC2808, at ${\rm T_{\rm eff}}$ below $\sim$12000, that marks the onset of radiative 
levitation effects (see discussion below). A similar result has been found for M~4 by Marino et al.~(2011),  
the general 
trend being that the bluer HB stars are on average more Na-rich and O-poor. 
As a consequence of these empirical results, we verified by performing a number of tests (see Appendix A)
how adequate is the use of 'standard' $\alpha$-enhanced 
spectra to determine bolometric corrections in this temperature regime. 
We found that results obtained by using standard $\alpha$-enhanced bolometric corrections, 
are unaffected by the presence of the observed CNONa anti-correlations.

Regarding the effect of radiative levitation, 
spectroscopic investigations by Behr~(2003),  Moehler et al.~(2003), Fabbian et al.~(2005)  
have disclosed that - similar to the case of NGC2808 - in these three clusters  
HB stars hotter than $\sim$12000~K display surface abundances of metal species like  Fe, Cr and Ti 
around solar or higher, due to radiative levitation in the atmosphere. Also, the surface He abundances are depleted 
by a factor of ten or more as a consequence of gravitational settling.
Following the approach used in Paper~I, we mimic this effect using 'standard' $\alpha$-enhanced HB stellar models 
by applying bolometric corrections appropriate for [Fe/H]=0.0 
(and scaled-solar mixture) when ${\rm T_{\rm eff}}$ is above 12000~K.
This 
is of course a crude approximation, which we are forced to adopt because of the lack of 
extended grids of both HB stellar  evolution and atmosphere models with a large range of chemical compositions, 
that include consistently the effect of radiative levitation.  
The underlying assumptions are that {\it (i)} the HB model evolutionary lifetimes,
{\it (ii)} the evolution of ${\rm T_{\rm eff}}$ and {\it (iii)} bolometric luminosity, 
and {\it (iv)} the structure of the corresponding model atmospheres 
are not affected by the radiative levitation, and that the relevant bolometric corrections are determined  
mainly by the enhanced metals. 
With these assumptions the effect of radiative levitation on bolometric corrections 
makes $m_{F160BW}$  fainter by about 0.15 and $m_{F336W}$ and $m_{F555W}$
brighter by about 0.1 and 0.05 respectively.
Paper~I has shown how 
this approximation allowed us to recover 
the three different He-abundances inferred by the multimodal MS, along the blue HB of NGC2808. 
We are therefore 
confident that a similar procedure will be adequate to constrain the He distribution 
along the HB of these three clusters, that we are going to discuss separately in the following section.

\subsection{The synthetic HB simulations}

The theoretical analysis performed in this paper is based  
on synthetic HB calculations, performed as described in Paper~I.
The analysis of NGC2808 was simplified by the fact that the initial He content of the cluster
subpopulations is somewhat 'quantized', as shown by the observed trimodal MS (Piotto et al. 2007). 
In case of M13, M3 and M79 there is no apparent 'quantization' of the MS, and we need to consider  
the possibility of a continuous distribution of Y along the HB. \\
We stress here that in these cases the modeling of the Y variations along the
HB becomes more uncertain since calculations require more assumptions.
Therefore the main goal of this work is not to attempt an exact description
of the Y distribution, that would be affected by our assumptions, but to 
understand whether helium plays a role in shaping these very different HBs.    
UV filters are much better suited than optical ones 
to address this issue.

Our synthetic HB calculations require the specification of at least 
4 parameters, plus the cluster age, that we assume to be equal to 
12~Gyr (see Section 3). These four parameters are 
the minimum value of Y (${\rm Y_{min}}$), the range of He abundances (${\rm \Delta Y}$), the mean value of the 
mass lost along the RGB ${\rm \Delta M}$ -- that in first instance (and for simplicity) we assume 
to be the same for each Y  -- and the spread around this mean value 
(${\rm \sigma(\Delta M)}$). 
Throughout our analysis we will assume that Y varies according to either a uniform probability 
distribution, from ${\rm Y_{min}}$ to 
${\rm Y_{min}+\Delta Y}$, or a Gaussian one. 
In this latter case ${\rm \Delta Y}$ is the 
1$\sigma$ spread around a prescribed mean value. 
Also for the RGB total mass loss we will assume a Gaussian  
distribution around the mean value ${\rm \Delta M}$ or a uniform distribution with 
minimum value equal to  ${\rm \Delta M}$. 
The idea behind this type of simulations (see, e.g., Caloi \& D'Antona~2005) is that the colour extension 
of the HB is driven mainly by the variation of Y rather than mass loss efficiency. A  
higher initial Y implies a lower mass at the tip of the RGB (for a fixed cluster age) hence, for a 
fixed value of ${\rm \Delta M}$, a smaller mass along the HB and a bluer colour (the He-core mass is affected 
by the variation of Y to a smaller extent).
Our synthetic HB code first draws randomly a value of Y, and determines the initial mass of the star 
at the RGB tip (${\rm M_{TRGB}}$) from interpolation 
among the BaSTI isochrones of the prescribed age. The mass of the corresponding object evolving along the HB 
(${\rm M_{HB}}$) is then given by ${\rm M_{HB} = M_{TRGB} - \Delta M}$ plus a Gaussian random perturbation 
${\rm \sigma(\Delta M)}$ (or a uniform probability between ${\rm \Delta M}$ and ${\rm \Delta M}$+${\rm \sigma(\Delta M)}$). 
The WFPC2 magnitudes of the synthetic star are then determined according to 
its position along the HB track with appropriate mass and Y-- obtained by interpolation among the available
set of HB tracks -- after an evolutionary time t has been randomly extracted. 
We determined t assuming that stars reach the ZAHB at a constant rate. 
We employed a flat probability distribution 
ranging from zero to ${\rm t_{HB}}$, where ${\rm t_{HB}}$ denotes the time spent from the ZAHB to 
the He-burning shell ignition along the early AGB. 
The value of ${\rm t_{HB}}$ is set by the mass with the longest lifetime (the lowest masses for a given Y and Z). 
This implies that for some synthetic object the 
randomly selected value of t will be longer than its ${\rm t_{HB}}$ or, in other words, that they have already evolved 
to the next evolutionary stages. 
Finally, the derived synthetic magnitudes are perturbed with a Gaussian 1$\sigma$ error determined from the 
data reduction procedure.

We detail now the effects on these results of possible variations of age and chemical composition
within the uncertainties described in Section~3.  
The mass distribution along the HB (as derived in our analysis) does not depend on the cluster age,
but changing the cluster age changes the mass loss necessary to reproduce the HB mass distribution.
An increase of age of $1$ Gyr leads to a decrease of the total mass loss along the RGB by $\sim0.02 M_{\odot}$ (because of 
the decreased evolving mass). \\      
One can simulate the effect of a variation of [Fe/H], [$\alpha$/Fe] and/or their sum by considering 
 that at these metallicities
 scaled solar and $\alpha$-enhanced models (also for the HB) with the 
 same total global metallicity [M/H] are coincident (Salaris et al. 1993).
 We have therefore considered a change $\Delta [M/H]=+0.3$ dex that corresponds to changes
 of [Fe/H] or [$\alpha$/Fe] or [Fe/H]+[$\alpha$/Fe].
 We have two effects here. The first one is on the derived mass distribution (at fixed Y) along the HB.
 For the redder part of the HB we use the ($m_{F336W}$,$m_{F336W}-m_{F555W}$) colour. 
 At fixed colour in the horizontal part of
 the ($m_{F336W},m_{F336W}-m_{F555W}$) CMD, an increase $\Delta$[M/H]=0.3 decreases the derived value of ${\rm M_{HB}}$ by
 $\sim$0.04 $M_{\odot}$.
 For the blue tails we use the ($m_{F160BW},m_{F160BW}-m_{F555W}$) CMD. At fixed ($m_{F160BW}-m_{F555W}$)
 an increase $\Delta$[M/H]=0.3 decreases the derived value of ${\rm M_{HB}}$ by $\sim0.01$ $M_{\odot}$ .
 This means that if a larger mass loss is needed to explain the blue HB stars
 (compared to the mass loss needed to explain the redder
 population), an increase of [M/H] tends to reduce this mass loss difference.
 The second effect is on the RGB mass (hence on the absolute value of the RGB mass loss). An increase $\Delta$[M/H]=0.3
 causes an increase of ${\rm M_{RGB}}$ by $\sim0.025$ $M_{\odot}$.
 The distribution of Y along the observed HBs is basically unaffected by the precise age and [M/H], because it is based on
 the variation of magnitude (in $m_{F336W}$ and $m_{F160BW}$, depending on the star colour) with Y, 
 a differential property that is unaffected by the exact model metallicity and age.

\section{Comparison between theory and observations}

In this section we will discuss separately the theoretical analysis of our three selected clusters. 
We stress again that allowing Y to vary within individual clusters, increases the number 
of free parameters needed to match the observed HB star distribution in the CMD. 
Leaving aside for a moment the open problem of the RGB mass loss, 
the distribution of Y among stars fed onto the HB will depend on the convolution of 
the unknown (at least for these three clusters) initial Y distribution among MS stars, with  
the Y-dependent evolutionary timescales along MS and RGB predicted by stellar models. 
As discussed before, in absence of constraints from the MS, we assumed here as a first order approximation  
either Gaussian or uniform probability distribution for Y among HB stars. 
As a consequence, the derived distribution of Y and mass range (at fixed Y) are most probably 
just a 'reasonable' approximation to the real ones. A much more solid result is the total range of Y covered 
by the stars in each individual cluster, because -- thanks to our UV CMDs -- this is strongly constrained by 
the observed stellar magnitudes $m_{F160BW}$ (when available).

\begin{figure}
\includegraphics[width=84mm]{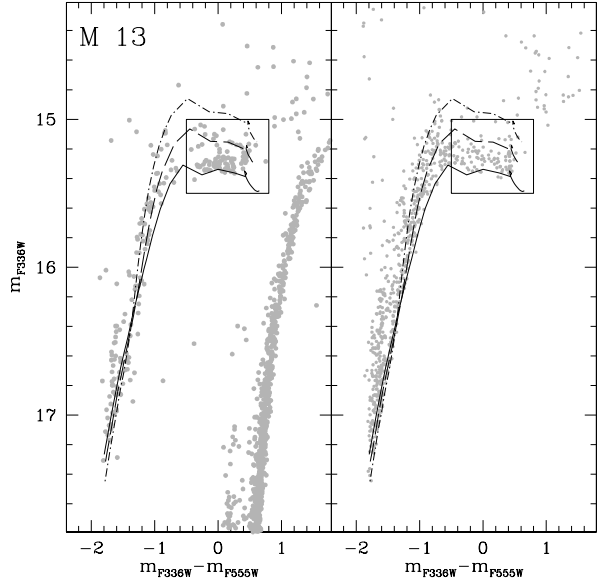}
\caption{Observed (left panel) vs synthetic (right panel)  ($m_{F336W},m_{F336W}-m_{F555W}$) CMD of M13 
HB. ZAHB sequences for Y=0.246 (solid line), 0.300 (dashed line) and 0.350 (dot-dashed line)   
are also displayed. The rectangular box marks the RHB region (see text for details).}
\label{synthM13_1}
\end{figure}

\subsection{M13}

Spectroscopic analyses of this cluster reveal a well developed Na-O anti-correlation 
(Sneden et al.~2004, Carretta et al.~2009), comparable to the case of NGC2808, 
but photometric analyses currently available do not show evidences for a multimodal MS. 
Here we analyze our HB photometry with -- as stated before -- the main aim of constraining the range of initial
He values 
of the cluster stars. The use of the F160BW filter -- not considered in the 
investigations by Caloi \& D'Antona~(2005) and D'Antona \& Caloi~(2008) -- will set strong constraints 
on the maximum value of Y. 

The first step of our analysis consists in defining the highest 
possible Y value necessary to fit the observed HB. To this purpose we 
calculated a synthetic HB with this preliminary set of parameters (uniform distributions for Y 
and Gaussian for M ):
 ${\rm Y_{min}}$=0.246, ${\rm \Delta Y}$=0.104, 
${\rm \Delta M}$=0.21${\rm M_{\odot}}$, ${\rm \sigma(\Delta M)}$=0.025${\rm M_{\odot}}$. 
Even though this selection of parameters is not optimized for a perfect fit 
to the observed stellar distribution along the HB, it will suffice to put a first general constraint to the maximum possible 
range of initial Y. The number of stars in this simulation is higher than the observed sample, to reduce the 
effect of statistical number fluctuations on the synthetic HB.

The parameter ${\rm \Delta Y}$ has been arbitrarily fixed to reach the maximum He mass fraction Y=0.35.
Once ${\rm Y_{min}}$ is chosen, the mean RGB total mass loss ${\rm \Delta M}$ is  
constrained by matching the red boundary of the observed HB (the subpopulation with the lowest Y 
has the largest HB mass at fixed age and ${\rm \Delta M}$). The value of ${\rm \sigma(\Delta M)}$ then introduces 
a general dispersion in the relationship between HB colour and local He abundance. The smaller the value of 
this parameter, the narrower the range of Y in a given colour bin along the HB. 
We notice that a very small value of 
${\rm \sigma(\Delta M)}$ (${\rm \sigma(\Delta M)}$=0.005 for example), would produce a sloped HB (brighter 
towards bluer colours because of a very tight colour-Y relationship) in the ($m_{F336W}$, $m_{F336W}-m_{F555W}$) CMD 
at colour ($m_{F336W}-m_{F555W}$)$> -$ 0.5, instead of the observed roughly horizontal structure (see Figure~\ref{synthM13_1}).

\begin{figure}
\includegraphics[width=84mm]{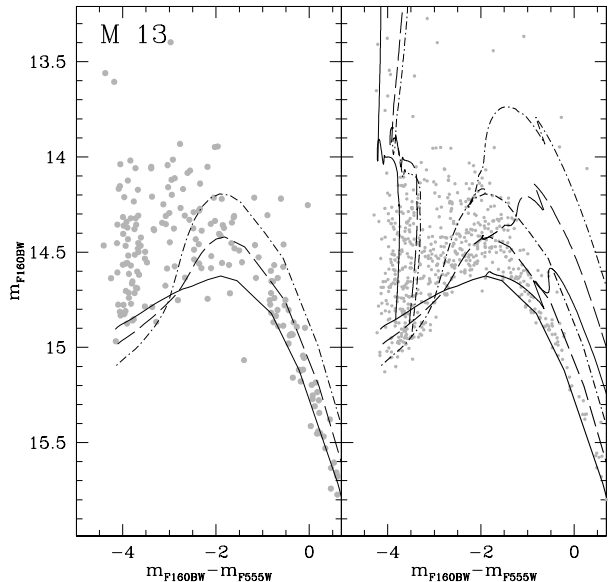}
\caption{As in Figure~\ref{synthM13_1} but for the ($m_{F160BW},m_{F160BW}-m_{F555W}$) CMD (see text for details).
In the right panel we also show two selected HB tracks corresponding to $M\sim0.57M_{\odot}$ and $M\sim0.50M_{\odot}$ 
for each ZAHB. }
\label{synthM13_2}
\end{figure}

As done in Paper~I, we determined the cluster distance modulus 
by fitting the synthetic data to the observed Red HB (RHB) (defined here 
as the region of the HB contained in the 
square box plotted in Figure~\ref{synthM13_1}) in the ($m_{F336W}, m_{F336W}-m_{F555W}$) plane. 
Notice how the RHB is roughly horizontal in 
this CMD, and allows us to estimate straightforwardly the cluster distance by comparing the synthetic data to 
the observed number distribution of objects in $m_{F336W}$, assuming a reddening 
E(B$-$V)=0.02 (see for example Ferraro et al. 1999). A fit to both the lower envelope -- defined as the faintest 
magnitude bin where the star counts are at least 2$\sigma$ above zero -- of the 
observed binned distribution in $m_{F336W}$, and the mean value of the $m_{F336W}$ RHB magnitudes ($<m_{F336W}>$=15.26) 
provides ${\rm (m-M)_0}$=14.32$\pm$0.05\footnote{The error bar is equal to half the width of the adopted  
bin size}, in fair agreement within the uncertainties with Ferraro et al. (1999; $(m-M)_0=14.43$).
The left and right panels of Figure~\ref{synthM13_1} show the observed and simulated 
($m_{F336W}, m_{F336W}-m_{F555W}$) CMD, respectively. We have also over-imposed three ZAHBs for Y=0.246 (solid line), 
0.300 (long dashed line) and 0.350 (dot-dashed line). 
Figure~\ref{synthM13_2} is equivalent to Figure~\ref{synthM13_1}, but for the ($m_{F160BW}, m_{F160BW}-m_{F555W}$) CMD.
Whilst in the 
($m_{F336W}, m_{F555W}$) CMD ZAHB sequences and synthetic stars with differing initial Y are largely degenerate along the 
bluest HB tail as a consequence of the huge increase of the bolometric correction in the optical filters, 
this same tail appears as a roughly horizontal sequence at ($m_{F160BW}-m_{F555W}$)$< -1.0$  
in the ($m_{F160BW},m_{F160BW}-m_{F555W}$) CMD. Theoretical ZAHB sequences display here different shapes at varying Y 
(see also Pietrinferni et al.~2006). It is clear from a comparison of the observed and synthetic HB that 
Y=0.35 for the bluest stars is too high. The lower envelope of the observed 
stellar distribution lies around the Y=0.30 ZAHB.
This implies that the maximum Y (${\rm Y_{max}}$) is $\sim$0.30.

As a conclusion, in the hypothesis that the colour extension of the HB is due to a range of initial Y values, 
${\rm Y_{min}}$=0.246 at the red end of the observed HB 
implies ${\rm Y_{max}}\sim$0.30 at the blue end,  i.e. the initial He-abundance spans a range ${\rm \Delta Y}\sim$0.05.
This constraint on ${\rm Y_{max}}$ is very general; the critical discriminating factor 
is provided by the $m_{F160BW}$ magnitude of the ZAHB at the blue end of the HB.

\begin{figure}
\includegraphics[width=84mm]{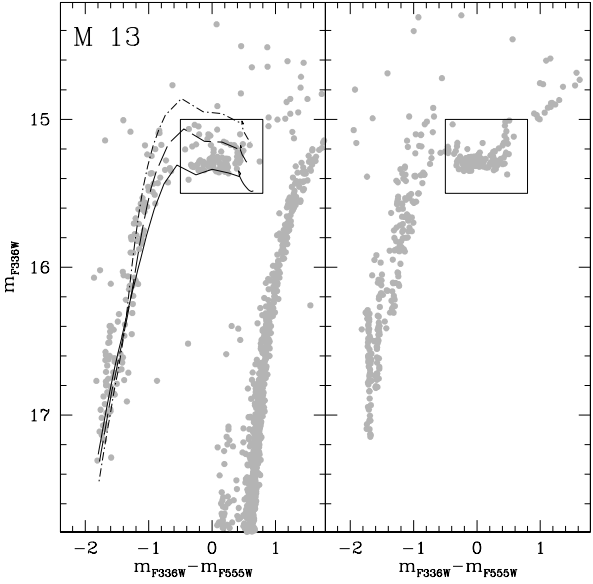}
\caption{As in Figure~\ref{synthM13_1} but for one realization of the best fit synthetic HB model discussed in the text.}
\label{synthM13_3}
\end{figure}

One can of course try a more accurate fit to the stellar distribution along the observed HB by comparing theoretical and
observed histograms 
of star counts as a function of colour and magnitude.  In this case 
a fine tuned calibration of the set of free parameters is necessary. In the assumption (for simplicity) 
of either uniform or Gaussian probability 
distributions for the mass loss and initial Y, one needs to split the simulation into three 
different sections, because no single continuous distribution of Y and mass allows to match the observed CMDs.
As a general method - applied also to M3 and M79 -, for each choice of the free parameters we have produced 
100 synthetic distributions, 
each of them with the same number of objects as in the observational sample. From this ensemble of simulations 
we have determined the mean value and associated error for each colour or magnitude bin used in their comparison with data. 


The star counts along the RHB on the red side of the gap at ($m_{F336W}-m_{F555W}$)$\sim -$ 0.5 
are best fit with ${\rm Y_{min}}$=0.246, ${\rm \Delta Y}$=0.01 and 
uniform probability distribution 
for Y, ${\rm \Delta M}$=0.21${\rm M_{\odot}}$, with a Gaussian spread ${\rm \sigma(\Delta M)}$=0.01${\rm M_{\odot}}$.  
The resulting distance modulus is ${\rm (m-M)_0}$=14.30$\pm$0.05, quite similar to the value obtained from the 
simulation in Figure~\ref{synthM13_1}.

\begin{figure}
\includegraphics[width=84mm]{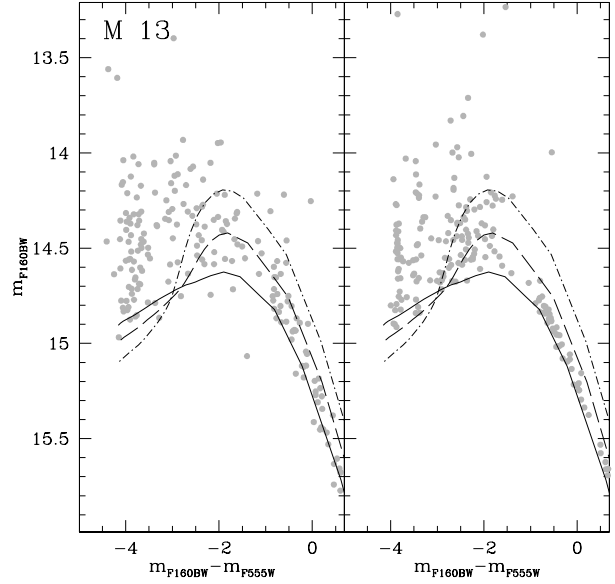}
\caption{As in Figure~\ref{synthM13_2} but for one realization of the best fit synthetic HB model discussed in the text.}
\label{synthM13_4}
\end{figure}

The second section of the simulation covers the region from the blue boundary of the gap 
to a colour ($m_{F160W}-m_{F555W}$)$\sim -$ 3.0 along the HB blue tail. In this case one needs a Gaussian distribution of Y, 
with mean value  ${\rm <Y>}$=0.285 and 1$\sigma$ spread ${\rm \Delta(Y)}$=0.012, mass loss 
${\rm \Delta M}$=0.235 ${\rm M_{\odot}}$ with a Gaussian spread ${\rm \sigma(\Delta M)}$=0.01${\rm M_{\odot}}$. 
The last section of the simulation covers the extreme tail of the HB, that is best reproduced with 
${\rm <Y>}$=0.300 with a Gaussian 1$\sigma$ spread ${\rm \Delta(Y)}$=0.003, and  
${\rm \Delta M}$=0.266 ${\rm M_{\odot}}$ with a Gaussian spread ${\rm \sigma(\Delta M)}$=0.002${\rm M_{\odot}}$. 
Figure~\ref{synthM13_3} and Figure~\ref{synthM13_4} display a comparison between the observed HB in the ($m_{F336W},m_{F336W}-m_{F555W}$)
 and ($m_{F160BW},m_{F160BW}-m_{F555W}$) CMDs, and one realization 
of the simulations with the parameter choice described before.
From the parameters of this simulations, it is clear that 
a good match to the observed star counts can be obtained only by relaxing the assumption of constant ${\rm \Delta M}$ for 
all initial He contents. In particular, it appears necessary to slightly increase on average ${\rm \Delta M}$ when Y increases, as 
already highlighted in Paper~I for the case of NGC2808. However it is important to note that,
the maximum Y range is fully constrained independently of the assumptions on mass loss.

Note that the slight increase of ${\rm \Delta M}$ with ${\rm Y}$ could be in principle due to an age variation at
increasing ${\rm Y}$.
In this scenario the variation of ${\rm \Delta M}$ corresponds to a variation of 
${\rm M_{TRGB}}$ with Y, keeping the total RGB mass loss unchanged.
However, these mass differences imply 
an increase of age with increasing Y (lower mass evolving along the RGB) of the order of $\sim 4Gyr$
that is hard to justify. In fact such a large age difference would be detectable from the cluster TO
 and sub-giant branch regions, at odds with the empirical evidence (Figure~\ref{m13m79} and Figure~\ref{m13m3}).

The mean He abundance along the HB of this simulation is ${\rm <Y>}$=0.270, with a 1$\sigma$ spread ${\rm \sigma(Y)}$=0.02. Notice that 
Salaris et al.~(2004) determined for this cluster Y=0.285$\pm$0.024 from the R-parameter, consistent with the results of 
the simulation.

We finally address the question of what happens if ${\rm Y_{min}}$ in RHB stars is increased. As a test, we have first considered 
a constant Y=0.280 for the RHB -- as proposed by Caloi \& D'Antona~(2005) and D'Antona \& Caloi~(2008) -- 
and produced the corresponding synthetic population by tuning appropriately the value of the mass range 
along the RHB. The distance modulus derived from the $m_{F336W}$ magnitudes of RHB stars is 0.14~mag larger than the case of our 
simulation with ${\rm Y_{min}}$=0.246. Given that the $m_{F160BW}$ magnitude along the extreme 
blue tail of the observed HB increases with increasing Y (see Figure~\ref{synthM13_2}), this higher distance modulus 
would imply Y$\sim$0.25 along 
the extreme HB (($m_{F160BW}-m_{F555W}$) $< -$3.5), i.e. a value lower than along the RHB,   
a result justifiable only assuming a strong anti-correlation between mass loss and Y.\\ 
This test also implies that in principle, for a fine tuned  ${\rm Y_{min}}$  intermediate between 0.246 and 0.280, the whole HB 
of M13 can be also modeled -- at least from the point of view of the magnitude levels in $m_{F336W}$ and $m_{F160BW}$ --  
employing a single value of Y. We find that 
when ${\rm Y_{min}}\sim$0.265, the ($m_{F336W},m_{F336W}-m_{F555W}$) and ($m_{F160BW},m_{F160BW}-m_{F555W}$) observed CMDs and star counts 
can be reproduced with a constant He abundance, for a total mass spread of 
$\sim$0.14 ${\rm M_{\odot}}$. Even if we cannot discard this scenario on the basis of the
HB analysis, we consider it very unlikely. In fact it would mean that M13 lost completely its
first generation of standard He 
stars, while preserving most of the He-enriched population. This would be hardly compatible with results from 
N-body simulations (see for example  Decressin, Baumgardt \& Kroupa 2008).

\begin{figure}
\includegraphics[width=84mm]{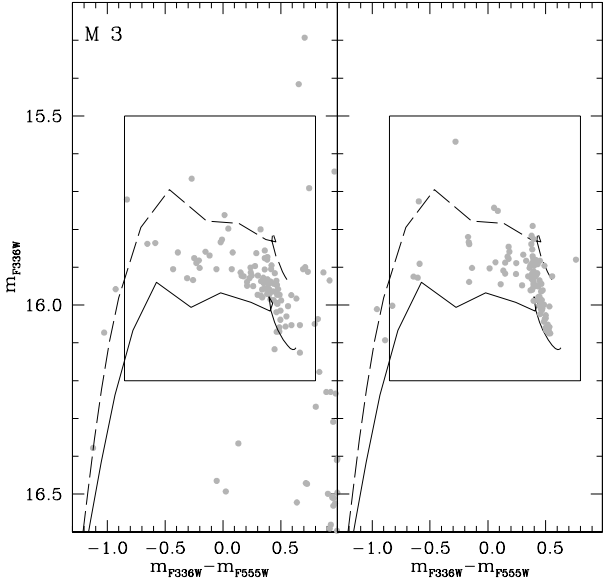}
\caption{Observed (left panel) vs one realization of the best fit synthetic (right panel) ($m_{F336W},m_{F336W}-m_{F555W}$) CMD of M3. 
The ZAHB sequences for Y=0.246 and Y=0.300 are also displayed. 
The rectangular box marks the region considered for the fit (see text for details).}
\label{synthM3_1}
\end{figure}

\subsection{M3}

Spectroscopic analyses of M3 RGB stars reveal a Na-O anti-correlation that is reduced in extension compared to M13 
(see, e.g., Sneden et al.~2004, Johnson et al.~2005 and references therein). Also in this case, the currently 
available photometric data do not show evidence of a multimodal MS.
Also, as well known, the HB of M3 displays a much shorter colour extension compared to M13. 
Here we discuss its properties only in the ($m_{F336W},m_{F336W}-m_{F555W}$) CMD, because no 
$F160BW$ images are available for this cluster.
Given the way in which HB stars are distributed in all CMDs involving $F255W$, it doesn't
give any additional information respect to the optical plane. Also given the relatively
short extension 
of the HB of M3, we decided not to use the shorter wavelength $F255W$ filter.\\
We considered in our analysis all objects within the rectangular box in 
Figure~\ref{synthM3_1}. Our synthetic calculations show that all these stars have ${\rm T_{\rm eff}}$ below the threshold 
for the onset of radiative levitation.  
There are three stars outside the box, that may be genuine HB objects, and appear spread along the 
bluest tail of the cluster HB, that is vertical also in this CMD. Given their extremely small number compared to the 
total sample of HB stars, and the fact that their location 
does not provide any further constraint on the cluster He distribution, we do not consider them in the 
theoretical analysis.

Following the same philosophy as for the case of M13, in the assumption of either uniform or Gaussian 
distributions for the mass loss and initial Y, observed star counts as a function of the 
$m_{F336W}$ magnitude and ($m_{F336W}-m_{F555W}$) colour are best reproduced with 
${\rm Y_{min}}$=0.246, ${\rm \Delta Y}$=0.02 and uniform probability distribution, a minimum RGB total mass loss   
${\rm \Delta M}$=0.122${\rm M_{\odot}}$ with a range (again uniform probability distribution) 
of values ${\rm \sigma(\Delta M)}$=9.0 $\times$ (Y-0.246) ${\rm M_{\odot}}$, where Y is the actual initial He mass fraction of 
the synthetic star. It is obvious that this dependence of ${\rm \sigma(\Delta M)}$ on Y  --  not imposed a priori 
but constrained by the fit to star counts as a function of magnitude and colour -- implies an increase of 
the mass loss spread with increasing Y, that is required by the observed shape of the CMD in the ($m_{F336W},m_{F336W}-m_{F555W}$) 
plane.

\begin{figure}
\includegraphics[width=84mm]{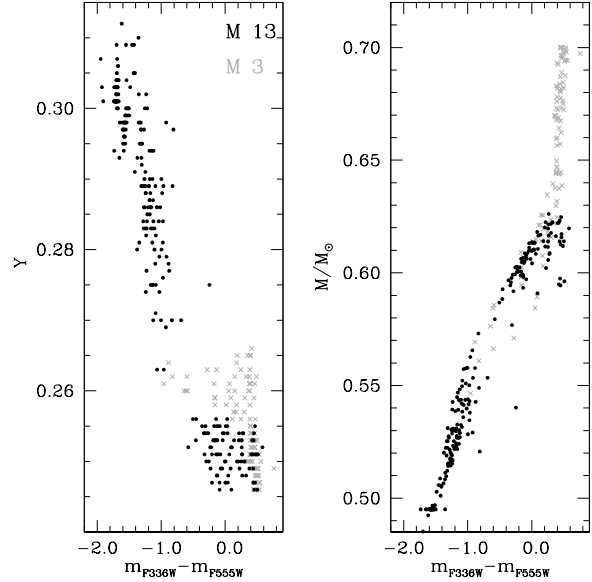}
\caption{Mass and Y distribution as a function of the ($m_{F336W}-m_{F555W}$) colour along the HB of M3 (grey crosses) 
and M13 (black dots).}
\label{synthM3_3}
\end{figure}

In fact as shown in Figure~\ref{synthM3_1}  the reddest part of the ZAHB describes a steeply sloped sequence, 
before 
turning horizontal at $m_{F336W}\sim16.0$. 
To match the observed star counts we need to increase the mass loss spread with increasing Y. 
This is necessary to reproduce both the steep 
increase of $m_{F336W}$ with decreasing colour when ($m_{F336W}-m_{F555W}$) is larger than $\sim$0.2 -- e.g., the lower 
He subpopulations have to cover a relatively small mass range --  and the almost constant $m_{F336W}$ 
when the colour ($m_{F336W}-m_{F555W}$) is bluer that this limit -- e.g., the higher He subpopulations 
have to cover an increasingly larger mass range. 
The mean He abundance obtained from the synthetic HB is ${\rm <Y>}$=0.256,
 with a 1$\sigma$ spread ${\rm \sigma(Y)}$=0.006, 
consistent with the value Y=0.253$\pm$0.013 obtained from the R-parameter by Cassisi et al.~(2003).
In the synthetic HB calculations we have disregarded all objects lying in the RRLyrae instability strip,
whose red and blue edges 
have been fixed at, respectively, log${\rm T_{\rm eff}}$=3.80 and log${\rm T_{\rm eff}}$=3.88 (see, e.g., the discussion in 
D'Antona \& Caloi~2008), which define a very narrow colour width of $\Delta(m_{F336W}-m_{F555W})<0.05$ at fixed $m_{F336W}$
approximatively in the range $0.35<(m_{F336W}-m_{F555W})<0.50$ .

The distance modulus obtained as described in the case of M13 
is ${\rm (m-M)_0}$=15.00$\pm$0.04 with standard reddening value E(B$-$V)=0.01 
which is fully in agreement with previous estimates (Ferraro et al. 1999; Harris 1996 -- catalogue version 2010).
In this case the maximum value of Y 
(${\rm Y_{max}}=0.266$) is essentially constrained by the brightest limit of the stellar distribution at  
($m_{F336W}-m_{F555W})\sim$0.2. 

Figure~\ref{synthM3_3} displays stellar mass and Y distributions as a function of the
($m_{F336W}-m_{F555W}$) colour in one 
representative synthetic realization of the cluster HB, compared to the corresponding best-fit composite simulation of 
M13. Not only the range of He values is obviously much larger in M13, but also the shape of the Y trend with colour 
is completely different in these two clusters. The mass distribution forms a very smooth continuous 
sequence, with the maximum and minimum masses along the HB distribution in M3 being appreciably larger than the 
corresponding values in M13.

A detailed comparison with recent results by Catelan et al (2009) and Caloi \& D'Antona (2008)
is reported in Appendix B.

\begin{figure}
\includegraphics[width=84mm]{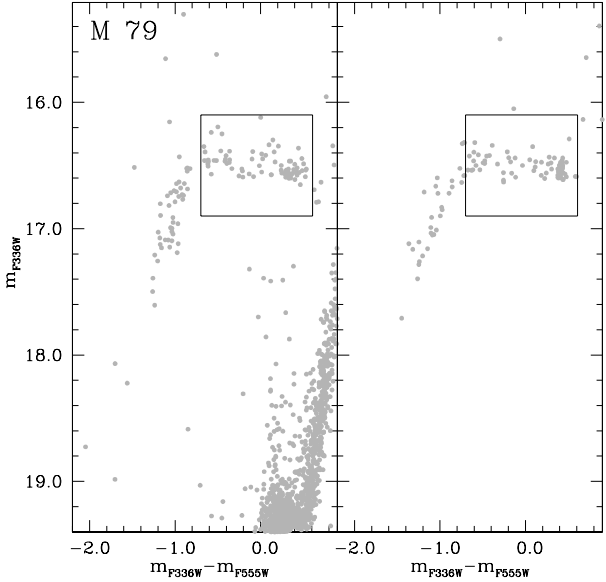}
\caption{Observed (left panel) vs one realization of the best fit synthetic (right panel) ($m_{F336W},m_{F336W}-m_{F555W}$) CMD of M79  
HB. The box marks the region used for the estimate of the cluster distance (see text for details).}
\label{synthM79_1}
\end{figure}

\subsection{M79}

\begin{figure}
\includegraphics[width=84mm]{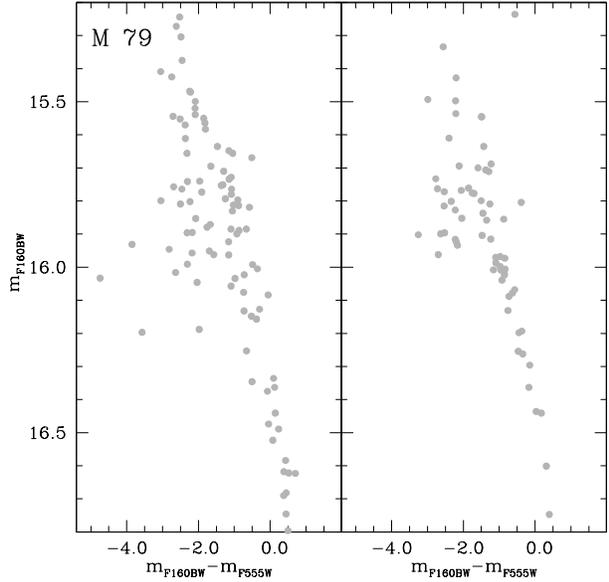}
\caption{As in Figure~\ref{synthM79_1} but for the ($m_{F160BW},m_{F160BW}-m_{F555W}$) CMD (see text for details).}
\label{synthM79_2}
\end{figure}

Spectroscopic analyses of M79 RGB stars disclose an extension of the 
Na-O anti-correlation that is intermediate between that of M13 and M3 
(see Sneden et al.~2004, Carretta et al.~2010), and photometric analyses do not reveal
evidences of a multimodal MS. Also the colour extension of its HB is 
intermediate between M13 and M3.
Following the same approach as for M13 and M3, we find that observed star counts 
as a function of magnitudes and colours are best reproduced with 
${\rm Y_{min}}$=0.246, ${\rm \Delta Y}$=0.035 and uniform probability distribution, a RGB total mass loss (at a given Y) ranging from  
${\rm \Delta M}$=0.11${\rm M_{\odot}}$ to ${\rm \Delta M}$=0.27${\rm M_{\odot}}$, with a uniform probability distribution.  
One realization of the best fit simulation is displayed in Figures~\ref{synthM79_1} and ~\ref{synthM79_2}.

The distance modulus has been determined from the fit to the stars contained within the box displayed in 
Figure~\ref{synthM79_1} (denoted here as RHB) 
as described for M13. 
Assuming E(B$-$V)=0.01 (Harris~1996)
we have obtained ${\rm (m-M)_0}$=15.64$\pm$0.05, values well in agreement 
with  both Ferraro et al. (1999) and Harris (1996).  For the RHB stars we determine 
an average ${\rm <Y>}$=0.263, with a 1$\sigma$ spread ${\rm \sigma (Y})$=0.010. The whole HB is characterized by 
${\rm <Y>}$=0.265, with a 1$\sigma$ spread ${\rm \sigma (Y)}$=0.010. This means that the full range of Y values is already present in 
the RHB sample. In the reasonable assumption that the spread of initial Y scales with the extension of the Na-O anti-correlation, 
we should expect to find among RHB stars the full range of Na-O abundances observed along the RGB. 
A range of He abundances is necessary along the RHB to match the observed thickness in $m_{F336W}$, and the 
large range of ${\rm \Delta M}$ (at fixed Y) is required to match the extended and almost horizontal structure of the RHB 
in the ($m_{F336W},m_{F336}-m_{F555W}$) CMD.

\begin{figure}
\includegraphics[width=84mm]{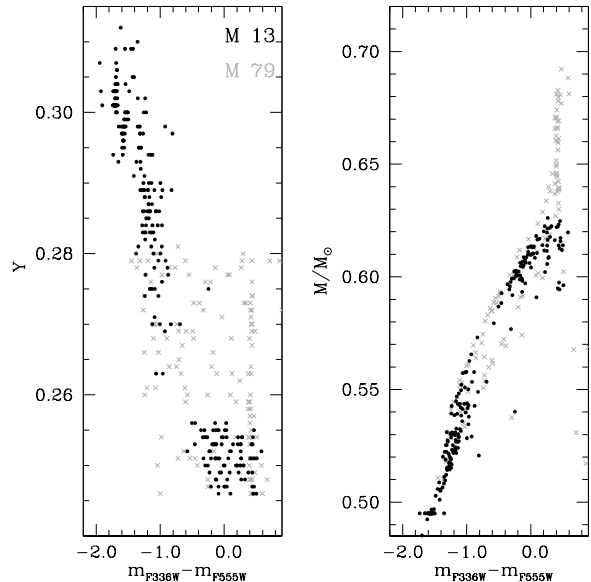}
\caption{Mass and Y distribution as a function of the ($m_{F336W}-m_{F555W}$) colour along the HB of M79 
(grey crosses) and M13 (black dots).}
\label{synthM79_5}
\end{figure}

Figure~\ref{synthM79_5} displays the mass and Y distribution as a function of the ($m_{F336W}-m_{F555W}$) colour in one 
representative synthetic realization of the cluster HB, compared to the M13 counterpart. 
The range of Y and the shape of the trend with colour are very different. The maximum mass 
of HB stars in M79 reaches higher values compared to M13, and the minimum HB mass is also higher than in M13.


\begin{table}
\caption{Filters adopted, number of observed HB and RRLyrae stars and values of ${\rm Y_{max}}$ 
and ${\rm <Y>}$ for the three clusters of our analysis.}
\begin{tabular}{@{}|p{1.1cm}{c}||p{0.7cm}{c}||p{1.0cm}{c}||p{0.6cm}{c}||p{0.6cm}{c}||p{0.5cm}{c}|}
\hline \hline
 CLUSTER  &  FILTER & ${\rm N_{HB}}$  &   ${\rm N_{RRLyrae}}$    & ${\rm Y_{max}}$  & ${\rm <Y>}$    \\ 
\hline
M3        & F336W  &   158   &  59 & 0.266 & 0.256 \\  
          & F555W  &  	&    &       &        \\
M13       & F160BW &   228   &    7 & 0.300 & 0.270\\ 
          & F336W  &         &    &	    &	     \\
	  & F555W  &        &    &	   &	    \\       
M79       & F160BW &   130   &   5 & 0.281 & 0.265 \\ 
          & F336W  &        &     &	    &	     \\
	  & F555W  &        &    &	   &	    \\
\hline
\hline
\label{tab:tot}
\end{tabular}
\end{table}

\section{Summary and conclusions}

There is a general consensus that the first parameter shaping the globular cluster HB morphology is
metallicity and the second one has been suggested by several authors (Lee et al. 2002; G10; D10)
to be the age.
While in general this picture may be surely adequate, however it is able to explain only few
young clusters in the Galaxy, while it fails to reproduce a number of \quotes{classical} cases 
where an additional ingredient (sometime called third parameter) is required.
The \quotes{third parameter} has been  proposed to be 
the stellar density or luminosity density (Fusi Pecci et al. 1993; D10) or the initial helium abundance (G10).\\    
In this framework we considered three GGCs with similar age and metallicity, M3,
M13 and M79, with the aim of studying their HBs. 
According to G10, M3 is younger by $\sim$ 2Gyr than M13, however we find, in agreement with 
a number of authors (Salaris \& Weiss 2002; 
Marin-Franch et al. 2009; D10), that these clusters differ in age by less than $1~Gyr$.

This triplet is well known to have extremely different HBs, with M13 displaying by far the
bluest morphology, and showing also evidence of at least 2 gaps at different effective
temperatures, while M3 has been proposed over the years as a prototype of a \quotes{normal} HB.\\ 
As done in Paper~I, we compared HST UV and optical CMDs 
with theoretical HB models, with the aim of understanding the impact (if any) 
of cluster-to-cluster Y variations on HB morphologies. 
At odds with the case of NGC2808, the three GCs analyzed in this work, do not show 
(so far) evidences of quantized MS, so our analysis has been performed without any
a priori knowledge of the Y distribution. As explained in Section~4, this forced us 
to enlarge the parameters space: the quantities needed are the minimum value of Y ($Y_{min}$) and its 
full range (${\rm \Delta Y}$), the mean value of the mass lost along the RGB $\Delta M$ and the dispersion 
around the mean value. As a consequence, it is hard in these cases to determine uniquely 
a full and detailed
representation of the Y variations along the observed HBs. A more stringent and robust derivation involves 
instead the largest value of Y (${\rm Y_{max}}$) that is clearly constrained
by the distribution of stars in the $(m_{F160BW};m_{F160BW}-m_{F555W})$ CMDs. 
For the case of M3, given the shorter
extension of its HB, the magnitudes $m_{F336W}$ give already a strong constraint on $Y_{max}$.\\

With these caveat in mind, we performed a detailed analysis by comparing observed CMDs 
with a large set of synthetic CMDs and 
ZAHBs from the BaSTI database (Pietrinferni et al. 2006). The effects of levitation for stars with 
$T_{\rm eff}>12000$ K and anti-correlations along the HBs (Marino et al. 2011; Gratton et al. 2011)
have been considered. \\
We find differences  ${\rm \Delta Y_{max}}\sim0.02-0.04$ between these GGCs. In particular M13 displays 
the largest value (${\rm Y_{max}}\sim 0.30$), M3 (${\rm Y_{max}}\sim 0.27$) the smallest one, and M79 is an
intermediate case with ${\rm Y_{max}}\sim 0.28$. These values are summarized in Table~1. They seem to qualitatively correlate with  
the differences in the temperature (colour) extensions of the cluster HBs. 
It is also interesting to note that our estimates of ${\rm Y_{max}}$ for these three clusters nicely
correlate with the observed range of light-element variations. In particular, M13 shows the most extreme 
Na-O anti-correlation (Sneden et al. 2004), with stars reaching [O/Fe]$=-1.1$ and [Na/Fe]$=0.7$, while 
M3 the least extended one, [O/Fe]$=-0.2$ and [Na/Fe]$=0.5$. This is in line with the strict
correspondence between HB colours and Na-O abundances observed in NGC2808  by Gratton et al. (2011).\\
The comparison between M13 and M3 is particularly interesting, because of several previous analyses of 
their HB morphology and Y distribution. For example Catelan (2009) performed a detailed analysis comparing the mean masses 
of the HBs of these clusters with age differences proposed by different authors. By using several 
mass-loss recipes, he found that there is no way to reproduce the different HBs of M3 and M13 only in terms of age,
but at least one additional ingredient should be invoked to account for the blue HB extension observed in M13.
Our best-fit model for M13 requires to split the simulation in three steps, given that no single continuous
distribution of Y and mass loss allows to match the observed CMDs. Thus we divided the HB in  
three groups: 1) the RHB with a typical ${\rm Y_{min}}=0.246$ and a mean mass lost = ${\rm \Delta M=0.21
M_{\odot}}$, 2) stars with $-3<(m_{F160BW}-m_{F555W})<-1.5$ best reproduced with ${\rm <Y>}=0.285$ and ${\rm \Delta
M=0.235M_{\odot}}$ and 3) stars with $(m_{F160BW}-m_{F555W})<-3$ which show a ${\rm <Y>}=0.30$ and ${\rm \Delta
M=0.266M_{\odot}}$. A fit with a population of stars with uniform $Y=0.265$ cannot be ruled out on the basis 
of only the HB analysis, while a single population of stars with $Y=0.28$, as proposed by Caloi \& D'Antona (2005) and 
D'Antona \& Caloi (2008), is incompatible with the distribution of stars in our UV CMD.\\
For M3 a single synthetic population with ${\rm Y_{min}}=0.246$ and distributed according to a uniform
probability distribution with ${\rm \Delta Y}=0.02$ is required. A total mass loss of ${\rm \Delta M
=0.122M_{\odot}}$ and a linear increase as a function of Y, as constrained by the fit of star
counts as a function of magnitude and colours, is needed. In the Appendix B we will discuss the Y distribution derived from 
our simulations in the context of 
the constraints posed by the analyses by Catelan et al. (2009) and D'Antona \& Caloi (2008); it turns out 
that our estimate of  ${\rm Y_{max}}=0.266$ is robust.\\
As highlighted by G10, while M13 seems to behave as other relatively massive clusters,
M3 appears to be peculiar and a more extended HB would have been expected in this case.
The age difference proposed by G10 between these two clusters would justify at least in part the different HB morphology,
in the framework in which age is the second parameter.
Our results would instead lead to think that M3 and M13 experienced a different amount of
enrichment of light elements. This would be compatible with the scenario proposed by G10 (see also
Carretta et al. 2009) that invokes a delayed cooling flow in the case of M3. 
In particular the HB simulations and derived ${\rm Y}$ distributions would suggest
that M13 is qualitatively similar to NGC2808, and that it likely experienced a similar star
formation, while M3 (and M79) probably had a less complex formation history. \\
Analyses based on a suitable combination of UV to optical photometry -- and when available observed
RR Lyrae period distributions --  and synthetic HB simulations
can provide not only insights on the HB second parameter problem, but in general 
they can potentially give some clues on the first stages of GCs formation and chemical evolution.
\\

We thank the referee Aaron Dotter for useful comments and suggestions that improved the presentation of this work.
This research is part of the project
COSMIC-LAB funded by the European Research Council (under contract
ERC-2010-AdG-267675).
SC acknowledges the financial support of 
the Ministero della Ricerca Scientifica e dell'Universita' PRIN MIUR 2007:
\lq{Multiple stellar populations in globular clusters}\rq\, ,
the Italian Theoretical Virtual Observatory Project and PRIN INAF 2011 \lq{Formation and Early Evolution of Massive Star Clusters}\rq\,

\newpage

{\bf Appendix A} \\
\\
{\bf The effect of anti-correlations on bolometric corrections} \\

We checked the impact of CNONa anti-correlations on the $\alpha$-enanched bolometric
corrections by performing a number of tests.
In particular, we have chosen three representative pairs of 
(${\rm T_{\rm eff}}$, $\log(g)$) values along our theoretical 
$\alpha$-enhanced ZAHB with [Fe/H]=$-$1.62, Y=0.246, namely  (5600~K,  2.40), (7000~K,  2.80), 
(10000~K,  3.50). For each of these pairs we have 
calculated a synthetic spectrum with the SYNTHE code by R.L. Kurucz (Sbordone et al. 2005) in the
wavelength range between 1200 and 7500 $\AA$, including all the atomic and molecular lines available
in the latest version of the Kurucz/Castelli linelist. 
The spectra were computed employing the LTE, plane-parallel
model atmospheres calculated with the ATLAS12 code (Castelli 2005).  
ATLAS12 adopts the \quotes{opacity sampling} method in the line opacity calculations, and  
allows to generate models for arbitrary chemical abundance mixtures.
All the model atmospheres and synthetic integrated spectra were computed with [Fe/H]=$-$1.62 and
Y=0.246 and both 
the standard [$\alpha$/Fe]=0.4 metal distribution used in the stellar evolution models, 
and the CNONa2 mixture by Sbordone et al.~(2011). This latter mixture displays -- compared to our reference 
$\alpha$-enhanced mixture --  enhancements of N and Na by 1.44~dex and 0.8~dex by mass, 
respectively, together with depletions of C and O by 0.6~dex and 0.8~dex, respectively, and 
unchanged CNO sum. This mixture is representative of extreme values of the CN and NaO anti-correlations 
observed in globular cluster stars (see Sbordone et al.~2011 for details). It is also very important to notice that low-mass 
stellar evolution models calculated for this mixture turn out to be identical to models calculated with the standard 
$\alpha$-enhanced metal distribution.

From the synthetic spectra we have calculated bolometric 
corrections to the F160BW, F336W and F555W filters used in our analysis, for both metal mixtures. 
Only at ${\rm T_{\rm eff}}=$5600~K there is a difference for the  
F336W filter (we did not consider the F160BW at this low 
temperature, because our photometry does not register cold HB stars in this passband) that amounts however to only 
0.03~mag (CNONa2 mixture stars appearing fainter). Given the small value of this difference and the fact 
that such extreme values of the anti-correlations appear not to be reached at such low temperatures along the HB, 
the outcome of our analysis of NGC2808 and the results of this paper, 
based on standard $\alpha$-enhanced bolometric corrections, are unaffected by the presence of the observed 
CNONa anti-correlations.

We have explicitly determined also the effect of He variations (at fixed metal content) on the bolometric corrections. 
We considered the point with (7000~K,  2.80) -- i.e. around the blue
edge of the RR Lyrae instability strip -- and calculated spectra for our standard $\alpha$-enhanced metal distribution, 
with Y=0.35. Compared to the case with standard Y=0.246, 
both F336W and F555W bolometric corrections are changed by less than 0.01~mag. The F160BW filter 
is affected  at this level by less than 0.02~mag. \\ \\ 


\begin{figure}
\includegraphics[width=84mm]{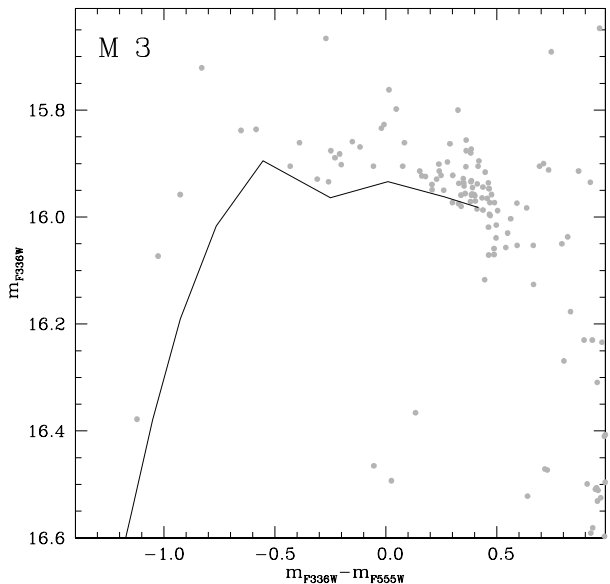}
\caption{Comparison of M3 HB with a ZAHB sequence with Y=0.256, for log(${\rm T_{\rm eff}}) >$ 3.89 (see text for details).}
\label{synthM3_4}
\end{figure}

\newpage
 
{\bf Appendix B} \\ 
\\
{\bf Comparison with previous studies on the HB morphology of M3.} \\

We address here two issues that arise from a comparison with recent results from the literature about the HB morphology of M3.
First of all, the range of Y derived from our simulations is about twice the value determined by Catelan et al.~(2009) 
from a comparison of theoretical ZAHB magnitudes and gravities with both Str\"omgren photometry, and
spectroscopically measured surface gravities and ${T_{\rm eff}}$. 
The size of the sample with spectroscopic gravities and ${T_{\rm eff}}$ estimates is 
relatively small, and within the corresponding error bars 
models with $Y\sim0.27$ cannot be easily ruled out. More compelling seems to be the constraint 
from the Str\"omgren CMDs. 
A fit of theoretical ZAHB models to the magnitude ($y$, $b$ and $v$) 
of the lower envelope of the observed HB at both the red side of the instability strip and 
at the 'knee' of the stellar distribution beyond the blue edge of the strip, before the sparsely populated and not very extended 
blue vertical tail -- where ZAHBs of different Y tend to become degenerate -- discloses a 
range of Y at most equal to $\Delta$Y$\sim$0.01, with 
Y increasing at the blue side of the strip. 
This disagreement with our results is however only apparent, for it can be very easily explained by recalling that 
to any given $Y$ value corresponds it is associated a dispersion in colour along the HB (see Section 4.1).
Figure~\ref{synthM3_4} illustrates clearly this point by 
comparing a theoretical ZAHB for Y=0.256 and ${\rm T_{\rm eff}}$ larger than the blue edge of the 
instability strip -- shifted by our derived distance modulus and assumed reddening --   
to the observed data.  
Notice how for ($0.2<(m_{F336W}-m_{F555W})<0.4$), that corresponds to the region of the 'knee'
in the Str\"omgren CMDs, 
this ZAHB fits nicely the observed lower envelope of the CMD, even though the average Y in that colour
range inferred by the best-fit synthetic model, is 0.260. Given the existence of a 
range of HB masses at fixed Y, this colour interval is also populated by stars with Y
 as low as $\sim$0.255, that are on average the faintest 
objects and hence determine the observed lower envelope of the magnitude distribution.

The second issue is related to the cluster rich RRLyrae population. 
The period distribution of fundamental RRLyrae is strongly peaked, and in the recent literature attempts have 
been made to reproduce both the peaked period distribution and the HB colour extension. Castellani, Castellani, \& Cassisi~(2005)
have shown that when keeping Y constant, a suitable bimodal mass distribution with two different mass dispersions 
is able to reproduce the observed features. On the other hand,  
D'Antona \& Caloi~(2008) assumed a range of Y and a very small mass dispersion for the mass loss along the RGB, 
and reproduced the observed HB colour extension and RRLyrae periods by calibrating the Y distribution (instead of 
the HB mass distribution).
It is interesting to notice that D'Antona \& Caloi~(2008) find from their simulations that 90\% 
of HB stars are expected to have Y$\le$0.265, in agreement 
with our results, 
and only 10\% of objects in their simulations reach Y values above our estimated maximum Y.

Here we have considered the synthetic objects in the best fit simulations that fall within the boundary of the instability strip 
(using the same ${\rm T_{\rm eff}}$ boundaries employed by D'Antona \& Caloi~2008) and were excluded from the comparison with the observed 
CMD. We have calculated their pulsational periods using the relationships by Di Criscienzo et al.~(2004) and compared with 
the observed distribution (Corwin \& Carney~2001). Our predicted period distribution does not reproduce the observed ones, which is probably 
not surprising given our 'simple' assumptions about the distribution of Y among stars fed onto the HB. 

\begin{figure}
\includegraphics[width=84mm]{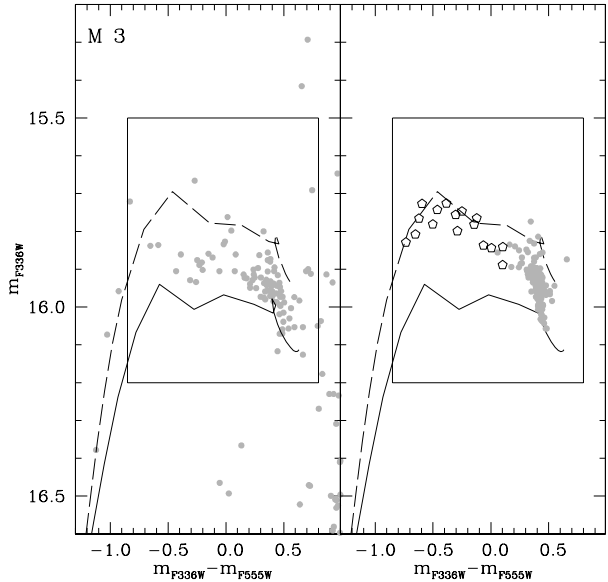}
\caption{Observed (left panel) vs one realization of a synthetic (right panel) ($m_{F336W},m_{F336W}-m_{F555W}$) CMD of M3 
using the Y distribution proposed by D'Antona \& Caloi~(2008). Open symbols denote synthetic objects 
with Y$>$0.270. The ZAHB sequences for Y=0.246 and Y=0.300 are also displayed. 
The rectangular box marks the region considered in the comparison (see text for details).}
\label{synthM3_5}
\end{figure}

As a test, we have calculated additional synthetic HB models for M3, employing this time 
the $ad hoc$ Y distribution\footnote{Our minimum Y 
value is 0.246 instead of 0.240} proposed by D'Antona \& Caloi~(2008 -- see their Figure~4) 
to reproduce both the observed period distribution in the instability strip and the $(B-V)$ colour distribution along the HB, while 
keeping the same mass loss for all HB stars.
As in  D'Antona \& Caloi~(2008) we employed a Gaussian mass distribution; our adopted mean 
mass loss is ${\rm \Delta M}$=0.160${\rm M_{\odot}}$ with 
a small (as discussed in D'Antona \& Caloi 2008) 
Gaussian spread ${\rm \sigma(\Delta M)}$=0.002${\rm M_{\odot}}$, and employed the same reddening and distance modulus as in the 
previous simulations. 
These choices allow to reproduce well 
the narrow period distribution (both mean value of the logarithm of the periods and the 1$\sigma$ dispersion) whereas 
a comparison of the the stellar distribution in the CMD at the blue side of the strip is less satisfactory. 
The left and right panels of Figure~\ref{synthM3_5} show the observed and simulated 
($m_{F336W}, m_{F336W}-m_{F555W}$) CMD. Notice how the observed shape of the CMD is reasonably reproduced as long as Y is below $\sim$0.270.  
For higher values of Y, synthetic stars tend to reach too bright magnitudes (while in the $BV$ plane they 
tend to be located along the more vertical blue part of the HB, at/beyond the 'knee'). This reinforces our estimate of 
the Y range as determined by  
our previous set of simulations -- that is the main purpose of our analysis -- 
and the need to increase the mass loss spread at the blue side of the instability strip -- instead of Y -- in order to match 
the CMD location of the bluer objects.


\newpage


       
\begin{thebibliography}{}



\bibitem[Bedin et al.(2004)]{2004ApJ...605L.125B} Bedin, L.~R., Piotto, G., 
Anderson, J., et al.\ 2004, ApJl, 605, L125 


\bibitem[Behr(2003)]{2003ApJS..149...67B} Behr, B.~B.\ 2003, ApJS, 149, 67 



\bibitem[Bragaglia et al.(2010)]{2010ApJ...720L..41B} Bragaglia, A., 
Carretta, E., Gratton, R.~G., et al.\ 2010, ApJl, 720, L41 

\bibitem[Buonanno et al. (1983)]{buon83} Buonanno, R., Buscema, G., Corsi,
C.~E., Ferraro, I., \& Iannicola, G.\ 1983, A\&A, 126, 278

\bibitem[Buonanno et 
al.(1985)]{1985A&A...145...97B} Buonanno, R., Corsi, C.~E., \& Fusi Pecci, F.\ 1985, A\&A, 145, 97


\bibitem[Busso et al.(2007)]{2007A&A...474..105B} Busso, G., et al.\ 2007, A\&A, 474, 105




\bibitem[Caloi \& D'Antona(2005)]{2005A&A...435..987C} Caloi, V., \& D'Antona, F.\ 2005, A\&A, 435, 987

\bibitem[Carretta et al.(2009)]{2009A&A...508..695C} Carretta, E., Bragaglia, A., Gratton, R., D'Orazi, V., \& Lucatello, S.\ 2009, A\&A, 508, 695 

\bibitem[Carretta et al.(2010)]{2010A&A...516A..55C} Carretta, E., Bragaglia, A., Gratton, R.~G., et al.\ 2010, A\&A, 516, A55 

\bibitem[]{} Castellani, M., Castellani, V., \& Cassisi, S.\ 2005, A\&A, 437, 1017

\bibitem[]{} Cassisi, S., Salaris, M., \& Irwin, A.W. 2003, ApJ, 588, 862



\bibitem[Castelli(2005)]{2005MSAIS...8...25C} Castelli, F.\ 2005, Memorie 
della Societa Astronomica Italiana Supplementi, 8, 25


\bibitem[Catelan(2009)]{2009Ap&SS.320..261C} Catelan, M.\ 2009, APS\&S, 320, 261 

\bibitem[Catelan et al.(2009)]{} 
Catelan, M., Grundhal, F., Sweigart, A.V., Valcarce, A.A.R., \& Cort{\'e}s, C.\ 2009, ApJl, 695, L97

\bibitem[Clement et al.(2001)]{2001AJ....122.2587C} Clement, C.~M., Muzzin, 
A., Dufton, Q., et al.\ 2001, AJ, 122, 2587 

\bibitem[Cohen \& Melendez(2005)]{2005AJ....129.1607C} Cohen, J.~G., \& Melendez, J.\ 2005, AJ, 129, 1607

\bibitem[Conroy \& Spergel(2011)]{2011ApJ...726...36C} Conroy, C., \& Spergel, D.~N.\ 2011, ApJ, 726, 36 

\bibitem[]{} Corwin, T. M., \& Carney, B. W. 2001, AJ, 122, 3183 

\bibitem[Dalessandro et al.(2008)]{2008ApJ...677.1069D} Dalessandro, E., 
Lanzoni, B., Ferraro, F.~R., Rood, R.~T., Milone, A., Piotto, G., 
\& Valenti, E.\ 2008, ApJ, 677, 1069 


\bibitem[Dalessandro et al.(2011)]{2011MNRAS.410..694D} Dalessandro, E., 
Salaris, M., Ferraro, F.~R., et al.\ 2011, MNRAS, 410, 694 

\bibitem[\protect\citeauthoryear{Dalessandro et 
al.}{2012}]{2012AJ....144..126D} Dalessandro E., Schiavon R.~P., Rood 
R.~T., Ferraro F.~R., Sohn S.~T., Lanzoni B., O'Connell R.~W., 2012, AJ, 
144, 126 




\bibitem[]{} D'Antona, F., \& Caloi, V.\ 2008, MNRAS, 390, 693


\bibitem[Decressin et al.(2007)]{2007A&A...464.1029D} Decressin, T., Meynet, G., Charbonnel, C., Prantzos, N., \& Ekstr{\"o}m, S.\ 2007, A\&A, 464, 1029

\bibitem[Decressin et al.(2008)]{2008A&A...492..101D} Decressin, T., Baumgardt, H., \& Kroupa, P.\ 2008, A\&A, 492, 101 

\bibitem[D'Ercole et al.(2008)]{2008MNRAS.391..825D} D'Ercole, A., 
Vesperini, E., D'Antona, F., McMillan, S.~L.~W., 
\& Recchi, S.\ 2008, MNRAS, 391, 825 

\bibitem[]{}Di Criscienzo, M., Marconi, M., \& Caputo, F. 2004, ApJ, 612, 1092



\bibitem[Dolphin(2000)]{2000PASP..112.1397D} Dolphin, A.~E.\ 2000, PASP, 
112, 1397 

\bibitem[Dotter et al.(2010)]{2010ApJ...708..698D} Dotter, A., Sarajedini, 
A., Anderson, J., et al.\ 2010, ApJ, 708, 698 


\bibitem[Fabbian et al.(2005)]{2005A&A...434..235F} Fabbian, D., Recio-Blanco, A., Gratton, R.~G., \& Piotto, G.\ 2005, A\&A, 434, 235 

\bibitem[Ferraro et al.(1997)]{1997ApJ...484L.145F} Ferraro, F.~R., Paltrinieri, B., Fusi Pecci, F., et al.\ 1997, ApJl, 484, L145 


\bibitem[Ferraro et al.(1998)]{1998ApJ...500..311F} Ferraro, F.~R., 
Paltrinieri, B., Pecci, F.~F., Rood, R.~T., 
\& Dorman, B.\ 1998, ApJ, 500, 311 

\bibitem[Ferraro et al.(1999)]{1999AJ....118.1738F} Ferraro, F.~R., 
Messineo, M., Fusi Pecci, F., et al.\ 1999, AJ, 118, 1738 

\bibitem[Ferraro et al.(2003)]{2003ApJ...588..464F} Ferraro, F.~R., Sills, 
A., Rood, R.~T., Paltrinieri, B., \& Buonanno, R.\ 2003, ApJ, 588, 464 


\bibitem[Freeman \& Norris(1981)]{1981ARA&A..19..319F} Freeman, K.~C., \& Norris, J.\ 1981, ARA\&A, 19, 319 

\bibitem[Fusi Pecci et al.(1993)]{1993AJ....105.1145F} Fusi Pecci, F., Ferraro, F.~R., Bellazzini, M., et al.\ 1993, AJ, 105, 1145 



\bibitem[Gratton et al.(2010)]{2010A&A...517A..81G} Gratton, R.~G., Carretta, E., Bragaglia, A., Lucatello, S., \& D'Orazi, V.\ 2010, A\&A, 517, A81 

\bibitem[Gratton et al.(2011)]{2011A&A...534A.123G} Gratton, R.~G., Lucatello, S., Carretta, E., 
Bragaglia, A., D'Orazi, V., \& Momany, Y.~A.\ 2011,  A\&A, 534, A123 

\bibitem[Grundahl et al.(1999)]{1999ApJ...524..242G} Grundahl, F., Catelan, 
M., Landsman, W.~B., Stetson, P.~B., 
\& Andersen, M.~I.\ 1999, ApJ, 524, 242 


\bibitem[Harris (1996)]{har96} Harris, W.E. 1996, AJ, 112, 1487


\bibitem[Hoyle \& Schwarzschild(1955)]{Hoyle55} Hoyle, F., \& Schwarzschild, M.\ 1955, ApJ, 121, 776 

\bibitem[Johnson \& Bolte(1998)]{1998AJ....115..693J} Johnson, J.~A., \& Bolte, M.\ 1998, AJ, 115, 693 

\bibitem[Johnson et al.(2005)]{2005PASP..117.1308J} Johnson, C.~I., Kraft, 
R.~P., Pilachowski, C.~A., et al.\ 2005, PASP, 117, 1308 


\bibitem[Iben \& Rood(1970)]{Iben70} Iben, I., Jr., \& Rood, R.~T.\ 1970, ApJ, 161, 587 

\bibitem[Lanzoni et al.(2007)]{2007ApJ...668L.139L} Lanzoni, B., 
Dalessandro, E., Ferraro, F.~R., Miocchi, P., Valenti, E., 
\& Rood, R.~T.\ 2007, ApJl, 668, L139 


\bibitem[Lee et al.(1987)]{1987fbs..conf..137L} Lee, Y.-W., Demarque, P., 
\& Zinn, R.\ 1987, IAU Colloq.~95: Second Conference on Faint Blue Stars, 137 


\bibitem[Lee et al.(1988)]{1988csa..proc..149L} Lee, Y.-W., Demarque, P., 
\& Zinn, R.\ 1988, Calibration of Stellar ages, 149 

\bibitem[Lee et al.(1990)]{1990ApJ...350..155L} Lee, Y.-W., Demarque, P., 
\& Zinn, R.\ 1990, ApJ, 350, 155 

\bibitem[Lee et al.(1994)]{1994ApJ...423..248L} Lee, Y.-W., Demarque, P., 
\& Zinn, R.\ 1994, ApJ, 423, 248 

\bibitem[Lee et al.(2002)]{2002IAUS..207..110L} Lee, Y.-W., Lee, H.-C., 
Yoon, S.-J., Rey, S.-C., 
\& Chaboyer, B.\ 2002, Extragalactic Star Clusters, 207, 110 



\bibitem[Marino et al.(2011)]{} 
Marino, A. F., Villanova, S., Milone, A.P., Piotto, G., Lind, K., Geisler, D., Stetson, P.B.\ 2011, ApJl, 730, L16

\bibitem[Mar{\'{\i}}n-Franch et al.(2009)]{2009ApJ...694.1498M} 
Mar{\'{\i}}n-Franch, A., Aparicio, A., Piotto, G., et al.\ 2009, ApJ, 694, 1498 

\bibitem[Mar{\'{\i}}n-Franch et al.(2010)]{2010ApJ...714.1072M} 
Mar{\'{\i}}n-Franch, A., Cassisi, S., Aparicio, A., 
\& Pietrinferni, A.\ 2010, ApJ, 714, 1072 


\bibitem[Milone et al.(2010)]{2010ApJ...709.1183M} Milone, A.~P., Piotto, 
G., King, I.~R., et al.\ 2010, ApJ, 709, 1183 


\bibitem[Milone et al.(2012)]{2012ApJ...745...27M} Milone, A.~P., Marino, 
A.~F., Piotto, G., et al.\ 2012, ApJ, 745, 27


\bibitem[Moehler et al.(2003)]{2003A&A...405..135M} Moehler, S., Landsman, W.~B., Sweigart, A.~V., \& Grundahl, F.\ 2003, A\&A, 405, 135 





\bibitem[Pasquini et al.(2011)]{2011A&A...531A..35P} Pasquini, L., Mauas, P., K{\"a}ufl, H.~U., \& Cacciari, C.\ 2011, A\&A, 531, A35 

\bibitem[\protect\citeauthoryear{Percival 
\& Salaris}{2011}]{2011MNRAS.412.2445P} Percival S.~M., Salaris M., 2011, MNRAS, 412, 2445

\bibitem[Pietrinferni et al.(2006)]{2006ApJ...642..797P} Pietrinferni, A., 
Cassisi, S., Salaris, M., \& Castelli, F.\ 2006, ApJ, 642, 797 


\bibitem[]{} Piotto, G., Bedin, L. R., Anderson, J., King, I. R., Cassisi, S., Milone, A. P., Villanova, S., Pietrinferni, A., \& Renzini, A. 2007, ApJ, 661, L53

\bibitem[Rey et al.(2001)]{2001AJ....122.3219R} Rey, S.-C., Yoon, S.-J., 
Lee, Y.-W., Chaboyer, B., \& Sarajedini, A.\ 2001, AJ, 122, 3219 

\bibitem[Rich et al.(1997)]{1997ApJ...484L..25R} Rich, R.~M., et al.\ 1997,  ApJl, 484, L25

\bibitem[Rood(1973)]{1973ApJ...184..815R} Rood, R.~T.\ 1973, ApJ, 184, 815 

\bibitem[Rood et al.(2008)]{2008MmSAI..79..383R} Rood, R.~T., Beccari, G., 
Lanzoni, B., Ferraro, F.~R., Dalessandro, E., 
\& Schiavon, R.~P.\ 2008, Memorie della Societa Astronomica Italiana, 79, 383

\bibitem[Rosenberg et al.(1999)]{1999AJ....118.2306R} Rosenberg, A., 
Saviane, I., Piotto, G., \& Aparicio, A.\ 1999, AJ, 118, 2306
 

\bibitem[Salaris et al.(1993)]{1993ApJ...414..580S} Salaris, M., Chieffi, A., \& Straniero, O.\ 1993, ApJ, 414, 580 

\bibitem[Salaris \& Weiss(2002)]{2002A&A...388..492S} Salaris, M., \& Weiss, A.\ 2002, A\&A, 388, 492 


\bibitem[Salaris et al.(2004)]{2004A&A...420..911S} Salaris, M., Riello, M., Cassisi, S., \& Piotto, G.\ 2004, A\&A, 420, 911 



\bibitem[Sbordone(2005)]{2005MSAIS...8...61S} Sbordone, L.\ 2005, Memorie 
della Societa Astronomica Italiana Supplementi, 8, 61 

\bibitem[Sbordone et al.(2011)]{2011A&A...534A...9S} Sbordone, L., Salaris, M., Weiss, A., \& Cassisi, S.\ 2011, A\&A, 534, A9 

\bibitem[\protect\citeauthoryear{Schiavon et 
al.}{2004}]{2004ApJ...608L..33S} Schiavon R.~P., Rose J.~A., Courteau S., 
MacArthur L.~A., 2004, ApJ, 608, L33 

\bibitem[Searle \& Zinn(1978)]{1978ApJ...225..357S} Searle, L., \& Zinn, R.\ 1978, ApJ, 225, 357 

\bibitem[Sneden et al.(2004)]{2004AJ....127.2162S} Sneden, C., Kraft, 
R.~P., Guhathakurta, P., Peterson, R.~C., 
\& Fulbright, J.~P.\ 2004, AJ, 127, 2162 



\bibitem[\protect\citeauthoryear{Valcarce \& Catelan}{2011}]{2011A&A...533A.120V} Valcarce A.~A.~R., Catelan M., 2011, A\&A, 533, A120 

\bibitem[Ventura et al.(2002)]{2002A&A...393..215V} Ventura, P., D'Antona, F., \& Mazzitelli, I.\ 2002, A\&A, 393, 215 


\bibitem[Zinn(1985)]{1985ApJ...293..424Z} Zinn, R.\ 1985, ApJ, 293, 424

\end{thebibliography}
\end{document}